\documentclass[a4paper,twocolumn,10pt]{article}

\usepackage[margin=1.8cm,top=2.0cm,bottom=2.0cm]{geometry}
\usepackage[T1]{fontenc}
\usepackage[utf8]{inputenc}
\usepackage{lmodern}
\usepackage{amsmath,amssymb,amsfonts}
\usepackage{physics}
\usepackage{siunitx}
\usepackage{graphicx}
\usepackage{booktabs}
\usepackage{cite}
\usepackage{microtype}
\usepackage{authblk}
\usepackage{xcolor}
\usepackage{hyperref}

\AtBeginDocument{\RenewCommandCopy\qty\SI}
\DeclareSIUnit\dBm{dBm}

\hypersetup{
    colorlinks=true,
    linkcolor=blue!70!black,
    citecolor=blue!70!black,
    urlcolor=blue!70!black
}

\title{\textbf{\Large Hyperentanglement-enabled Quantum Privacy Amplification\\
for Classical-Quantum Coexistence in Fiber Networks}}

\author[1,2]{\textbf{Domenico Ribezzo}\thanks{domenico.ribezzo@unifi.it}}
\author[1]{\textbf{Giulia Guarda}}
\author[2]{\textbf{Alessandro Zavatta}}
\author[1,2]{\textbf{Davide Bacco}\thanks{davide.bacco@unifi.it}}

\affil[1]{\small Dipartimento di Fisica e Astronomia, Universit\`a di Firenze, 50125 Firenze, Italy}
\affil[2]{\small Istituto Nazionale di Ottica, CNR-INO, 50125 Firenze, Italy}

\date{}

\begin{document}

\maketitle

\begin{abstract}
Integrating quantum communication into existing fiber infrastructures requires coexisting with high-power classical traffic, which generates noise in the quantum channel. Spontaneous Raman Scattering (SpRS) is the dominant noise source, generating a broadband photon background that degrades quantum states and, in Quantum Key Distribution (QKD), can entirely suppress secure key generation. In this work, we implement single-copy entanglement distillation utilizing energy-time and frequency hyperentanglement, and we exploit it to perform Quantum Privacy Amplification (QPA) directly at a physical layer. Strict frequency anti-correlations of spontaneous parametric down-conversion are used to selectively discard uncorrelated Raman photons while preserving genuine entangled pairs, suppressing accidental coincidences by a factor $1/N$, where $N$ is the number of frequency channel pairs. Using passive borosilicate-glass photonic integrated circuits and standard DWDM demultiplexers over a 20~km single-mode fiber testbed, we demonstrate a 4~dB improvement in tolerance to classical laser launch power $P_C$ when QPA is performed, extending the secure key threshold from $P_C=-13.3$~dBm to $P_C=-9.3$~dBm. The result shows that off-the-shelf telecom components can be repurposed as physical-layer quantum privacy amplifiers.
\end{abstract}

\vspace{0.2cm}

\section{Introduction}

The vision of a global Quantum Internet~\cite{wehner2018quantum} relies on the ability to transmit quantum signals over existing optical fiber infrastructures, which are already extensively deployed worldwide and carry high-power classical data traffic. Sharing this infrastructure offers a practically viable path towards the large-scale deployment of quantum networks, but introduces a substantial technical challenge: quantum information is encoded at the single-photon level, while classical channels typically operate at milliwatt power levels~\cite{kumar2015coexistence, patel2012coexistence, wu2025integration}. As a consequence of this power disparity, roughly ten orders of magnitude, classical light continuously generates noise in the quantum channels due to in-fiber nonlinear processes.

Among these processes, Spontaneous Raman Scattering (SpRS) is particularly relevant to quantum–classical coexistence, as it generates broadband, stochastic photons that can spectrally overlap with the quantum channel and directly contribute to its noise~\cite{eraerds2010quantum}. Other nonlinear effects, such as Four-Wave Mixing (FWM) and Brillouin scattering, generally have more limited spectral bandwidths, making it easier to filter out the resulting noise from the quantum channel~\cite{agrawal2012nonlinear}. Furthermore, Kerr nonlinearity can affect quantum transmission through effects such as Self-Phase Modulation (SPM) and Cross-Phase Modulation (XPM)~\cite{agrawal2012nonlinear,shapiro2006single}. In particular, XPM can induce phase fluctuations through the interaction with the co-propagating classical channels~\cite{wegener2004effect}. In the following, we neglect phase-related effects and focus exclusively on the contribution of SpRS to the quantum-channel noise. In the implemented entanglement-based QKD scheme, these noise photons generate accidental coincidences in the receivers, raising the quantum bit error rate (QBER) and eventually suppressing the secure key rate (SKR) entirely.

Conventional mitigation strategies -- including temporal gating, ultra-narrow filtering, and counter-propagation -- reduce the noise at the cost of added loss, increased complexity, or infrastructure constraints. Multi-core fibers (MCFs) can provide physical isolation between quantum and classical modes~\cite{wu2025integration}, but remain unavailable in most deployed networks. High-dimensional quantum states in principle tolerate higher error rates~\cite{cozzolino2019high, zahidy2024practical, zhong2024hyperentanglement}, but demand correspondingly more complex measurement hardware. This motivates the investigation of whether the benefits of high-dimensional quantum resources can be harnessed while retaining the simplicity of conventional measurement schemes.

In this work, we show that noise robustness and measurement simplicity do not need to be mutually exclusive. Our approach relies on single-copy entanglement distillation~\cite{simon2002polarization, barreiro2005generation, ecker2021experimental}, a simplified version of entanglement distillation~\cite{bennett1996concentrating,bennett1996purification}, applied to photon pairs that are simultaneously entangled in both energy-time and frequency degrees of freedom (DoF). The frequency correlations allow genuine photon pairs to be distinguished from uncorrelated Raman-scattered noise, enabling noise rejection at the detection stage and improving entanglement in energy-time DoF, thereby reducing the QBER for QKD.

This mechanism is precisely what Deutsch \textit{et al.} called \textit{Quantum Privacy Amplification} (QPA)~\cite{deutsch1996quantum}: performing entanglement distillation before measuring the DoF encoding the key reduces the eavesdropper's correlations with the transmitted photons, restoring the capacity for secure key extraction in noise regimes where classical post-processing alone would fail. Recent implementations of QPA via polarization and energy-time hyperentanglement have demonstrated this principle~\cite{ecker2021experimental, sohr2024overcoming, hu2021long}. Our work extends it to a scenario of practical relevance: rather than using generic noise sources, we co-propagate the entangled photons together with a classical laser operating in the telecom C-band, in a fiber spool, reproducing the spectral noise due to SpRS encountered in deployed networks relying on wavelength division multiplexing. Moreover, frequency is a more practical DoF compared to polarization for fiber infrastructure due to its intrinsic stability, and can be easily manipulated with standard telecommunication components, such as off-the-shelf DWDM.

We use passive borosilicate-glass photonic integrated circuits (PICs) as Franson interferometers to generate and analyze energy-time entanglement from an SPDC crystal pumped with a continuous-wave laser, while standard ITU-T 100-GHz DWDM filters are used to generate frequency-bin entanglement; this way, we generate hyperentanglement in the energy-time and frequency degrees of freedom. In a 20~km single-mode fiber coexistence testbed, the protocol extends the maximum classical launch power compatible with secure key generation from \SI{-13.3}{\dBm} to \SI{-9.3}{\dBm}, a \SI{4}{\decibel} improvement in noise tolerance while keeping the detection hardware identical to that of a conventional entanglement-based QKD system.

\section{Quantum Privacy Amplification}
\label{sec:QPA}

The security of entanglement-based QKD relies on a fundamental property of quantum entanglement: if Alice and Bob share a pure Bell state $\ket{\Phi^+}$, their joint quantum state cannot exhibit correlations with any third party. In practice, channel noise and disturbances introduced by Eve, a potentially malicious eavesdropper, produce indistinguishable effects from a quantum physics perspective. These effects degrade the distributed quantum state, resulting in a mixed state described by the density matrix:
\begin{equation}
	\rho_{AB} = (1-p)\ketbra{\Phi^+}{\Phi^+} + p\,\rho_{\rm noise}.
	\label{eq:rho_AB}
\end{equation}
Here $\rho_{AB}$ is the density matrix of the joint state shared by Alice and Bob, $\ket{\Phi^+}$ is the target Bell state, $p\in[0,1]$ is the mixing parameter quantifying the contribution of the noisy component, and $\rho_{\rm noise}$ is the density matrix describing the corresponding noise state. As $p$ increases, the shared state progressively deviates from the ideal Bell state, leading to an increased QBER. Standard QKD post-processing algorithms operate in two sequential stages. First, in the information reconciliation stage, Alice and Bob correct discrepancies between their respective bit strings -- arising from channel noise or eavesdropping -- to obtain a shared, error-free key. Subsequently, the privacy amplification stage compresses this key using universal hash functions, shortening it so as to eliminate, down to a specified security level, any information Eve may have acquired during the transmission. This scheme guarantees the security of the final key only when the mutual information between Alice and Bob, $I(A:B)$, exceeds the mutual information between Alice and Eve, $I(A:E)$, which is a measure of the information accessible to Eve. When this condition holds, there exists a privacy amplification procedure capable of producing a secret key of positive length; otherwise, no secure key can be extracted from the raw data, regardless of the post-processing algorithm used. For the BBM92 protocol under collective attacks, the asymptotic secure key rate vanishes when the quantum bit error rate reaches $QBER \geq 11\%$ (equivalently, when the interference visibility drops to $V \leq 78\%$). This threshold, derived from the Shor–Preskill security proof applied to the entanglement-based BBM92 scheme \cite{shor2000simple}, represents a fundamental limit: beyond it, no amount of classical post-processing -- however sophisticated -- can extract a secret key, since Eve's potential information about the raw key equals or exceeds the mutual information between Alice and Bob.

QPA circumvents this limit by intervening at the quantum layer, prior to measurement. Rather than allowing degraded entangled states to reach the detectors, this approach first distills them into a smaller number of higher-fidelity pairs. By purifying $\rho_{AB}$ toward the maximally entangled state $\rho_{AB}^{ideal}=\ketbra{\Phi^+}{\Phi^+}$ via local operations and classical communication (LOOC), one reduces Eve's residual entanglement with the system before any measurement takes place, effectively lowering the QBER and reopening the security regime that classical post-processing alone could not access~\cite{bennett1996concentrating,bennett1996purification,sohr2024overcoming}.

\section{Experimental Implementation}
\subsection{Single-Copy Entanglement Distillation}

The original QPA proposal~\cite{deutsch1996quantum, bennett1996purification} uses two noisy entangled pairs to produce a single pair with higher fidelity. Each party applies a CNOT gate to their two photons, using one as the control and the other one as the target, and then measures the target photon. The idea behind this protocol is that the CNOT gate propagates a potential bit-flip error from the control qubit to the target qubit. If the control photon of Alice or Bob has flipped, their target measurements disagree, and the control pair is discarded. Likewise, if one of the target photons has flipped, the same disagreement occurs, but in this case a valid control pair is sacrificed. The error goes undetected only when two flips compensate for each other in the parity check. That is, when both photons of the same part (control and target) are flipped, or when the control on one side and the target on the other are flipped simultaneously. It is possible to show that, provided that the initial fidelity satisfies $F>0.5$, the output fidelity increases after distillation, since the probability of such simultaneous multi-photon errors occurring increases more slowly than the probability of a single bit-flip~\cite{bennett1996purification}.

However, in photonic implementations this method is generally inconvenient:  it requires simultaneous transmission of two independent pairs, whose probability scales as $\eta^4$, where  $\eta$ is the channel transmission, and a deterministic two-photon CNOT gate that cannot be implemented with passive linear optics~\cite{pan2001experimental,o2003demonstration,gasparoni2004}. Single-copy approaches overcome these limitations by processing individual photon pairs. Early implementations relied on local filtering on a single degree of freedom~\cite{horodecki2001distillation,kwiat2001experimental}; while effective at re-balancing non-maximally entangled pure states, this approach cannot purify mixed states resulting from phase noise or decoherence. Single-copy distillation based on hyperentanglement goes beyond these problems by encoding two copies of the Bell state in different degrees of freedom of the same photon pair. The CNOT-like gate is then a local operation on a single photon, which can be implemented deterministically with linear optics, and the protocol rate scales as $\eta^2$, rather than $\eta^4$~\cite{ecker2021experimental, hu2021long, sohr2024overcoming}.

\subsection{System Architecture}

\begin{figure*}[t]
	\centering
	\includegraphics[width=0.85\linewidth]{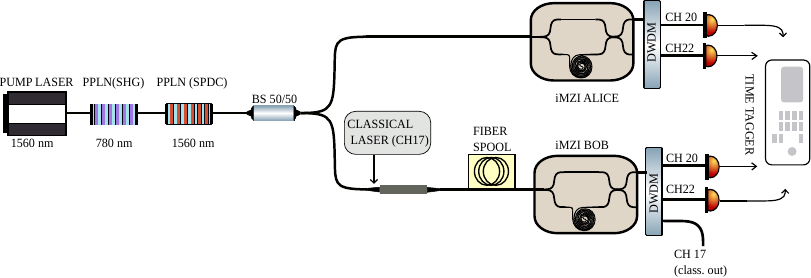}
	\caption{\textbf{Experimental setup.} Photon pairs hyperentangled in energy-time and frequency are generated via cascaded SHG and SPDC in PPLN waveguides and distributed to Alice and Bob. Each station analyzes energy-time correlations with a passive borosilicate-glass integrated imbalanced Mach-Zehnder interferometer (iMZI) in a Franson configuration. Bob's photon co-propagates over 20~km of standard SMF with a classical channel at DWDM CH~17, and DWDM demultiplexers on both sides select conjugate frequency-bin pairs to perform single-copy distillation.}
	\label{fig:setup}
\end{figure*}

The experimental architecture used in this experiment is shown in Fig.~\ref{fig:setup}. Photon pairs are generated by a cascade of two PPLN waveguides: a CW fiber laser at $\lambda_0 = 1560.48$~nm (DWDM CH~21, linewidth $\Delta\nu_p \approx 7.6$~kHz) first drives second-harmonic generation at 780.24~nm, which then pumps SPDC producing degenerate pairs centered at 1560.48~nm. Although this cascaded scheme does not offer any particular advantages over a single crystal pumped by a 780~nm laser, it does not introduce any significant drawbacks either, while making the setup easily reconfigurable for other purposes, such as homodyne measurements.

At this point, the entangled photon pairs directed to Alice and Bob are separated by a 50:50 beam splitter. This introduces a 3~dB loss: in half of the cases, both photons belonging to the same pair will be directed towards the same output and therefore discarded. However, it ensures that, when split, the signal and idler photons are randomly assigned to either Alice or Bob. This random assignment is essential for the frequency entanglement generated in the next stage.

After the beam splitter, Alice’s photon is directly sent to an imbalanced Mach–Zehnder interferometer (iMZI), a passive borosilicate-glass integrated photonic circuit. Here, the long arm introduces a delay of $\tau = 800ps$ with respect to the short arm. An identical interferometer is used for Bob. Before entering Bob’s interferometer, however, the photon propagates through a 20-km-long spool of single-mode fiber (SMF-28), showing 5.35~dB of attenuation. A classical optical signal at 1563.86 nm (ITU DWDM channel 17) is multiplexed onto the same fiber and co-propagates with the quantum signal, emulating an ongoing classical communication channel. The phases of the two iMZIs are thermally adjusted using a Peltier element placed beneath the PICs. This approach changes the overall temperature of the PIC rather than locally heating a specific section of the waveguide. Therefore, no active stabilization loop is required~\cite{guarda2024quantum}, simplifying the overall system. 

The configuration of Alice’s and Bob’s interferometers is the typical Franson interferometer scheme~\cite{franson1989bell}. The cross-correlation histogram of the detection events recorded after the two interferometers shows a characteristic three-peak structure. The two side peaks have a constant height, whereas the height of the central peak oscillates. The two side peaks correspond to the cases in which the photon entering Alice’s interferometer takes the long path while Bob’s photon takes the short path, or vice versa. These events are distinguishable based on their relative arrival time. In contrast, when both photons take the same path in their respective interferometers -- either both taking the long paths or both taking the short paths -- they arrive at the detectors at the same time. The two corresponding paths are therefore indistinguishable, since the pump driving the nonlinear photon-pair generation is continuous-wave. Hence, the paths cannot be determined from the photon generation and detection times.
Indistinguishable long–long and short–short probability amplitudes interfere, causing the central peak to oscillate. This phenomenon produces the characteristic Franson interference.
For this interference to occur, the interferometer delay $\tau$ must be longer than the coherence time of the individual photons $\tau_c^{\mathrm{ph}}$, so that single-photon interference is suppressed, but shorter than the coherence time of the pump $\tau_c^{\mathrm{pump}}$, which preserves the coherence between the two two-photon alternatives:
$$
\tau_c^{\mathrm{ph}} < \tau < \tau_c^{\mathrm{pump}}.$$

In our setup, the interferometer imbalance is $\tau = 800$ ps, satisfying these conditions.

Hence, the output is the maximally entangled Bell state $\ket{\psi_t}$:

\begin{equation}
	\ket{\psi_t} = \frac{1}{\sqrt{2}}\left(\ket{SS} + e^{i\phi_t}\ket{LL}\right)
	\label{eq:franson}.
\end{equation}
This state describes a two-dimensional entanglement in the energy-time degree of freedom. In order to perform single-copy entanglement distillation and using the QPA, we need to generate entanglement in an additional degree of freedom.

We exploit the fact that the signal and idler photons are randomly split between Alice's and Bob's paths and therefore have different central wavelengths. The outputs of each iMZI are sent to a demultiplexer. The demultiplexer divides the SPDC spectrum into an even number of frequency channels symmetrically distributed around its central peak. This spectral discretization provides access to the frequency entanglement in a discrete-variable encoding. Indeed, whenever Alice receives the signal photon, Bob receives the idler photon, and vice versa. Moreover, energy conservation imposes a fixed relation between their frequencies, which must sum to the pump frequency. The resulting state can therefore be written as:

\begin{equation}
    \ket{\psi_\omega} = \frac{1}{\sqrt{N}}\sum_{k=1}^{N}\ket{\omega_k}_A\ket{\omega_{-k}}_B.
\label{eq:state_freq}
\end{equation}

In our experiment, we use only one pair of 100-GHz DWDM demultiplexers, selecting the channels immediately to the left and right of the SPDC peak, namely channels 20 and 22. The channel passband of 100~GHz corresponds to $\approx$ 0.8~nm, with adjacent channel isolation > 35~dB. The frequency-entangled state can therefore be rewritten as\footnote{The choice of two frequency channels symmetric around the SPDC peak ensures that they carry approximately equal probability amplitude, since the
joint spectral amplitude is itself approximately symmetric around the	phase-matching peak, which justifies the equal-weight normalization $1/\sqrt{N}$ in Eq.~\eqref{eq:state_freq}.}:

\begin{equation}
    \ket{\psi_\omega} = \frac{1}{\sqrt{2}}\left(\ket{\omega_{20}}_A\ket{\omega_{22}}_B + e^{i\phi_{\omega}}\ket{\omega_{22}}_A\ket{\omega_{20}}_B\right).
	\label{eq:state_freq_2}
\end{equation}

This corresponds to a two-dimensional frequency-bin entangled state. We can now write the final state as the tensor product of the two two-dimensional entangled states, resulting in a four-dimensional hyperentangled state:

\begin{equation}
\ket{\Psi} = \frac{1}{2}\left(\ket{SS}+e^{i\phi_t}\ket{LL}\right)\otimes\left(\ket{\omega_1\omega_2} + e^{i\phi_{\omega}}\ket{\omega_2\omega_1}\right),
	\label{eq:hyper_state}
\end{equation}
where, for simplicity, we introduce a reduced notation for the frequency degree of freedom. Here, the entanglement is encoded simultaneously in the energy-time and frequency degrees of freedom, providing a four-dimensional hyperentangled state.

Finally, > 90\% efficiency superconducting nanowire single-photon detectors (SNSPDs) by Quantum Opus and an ultra-low jitter time tagger by Swabian Instruments (Timetagger X) are used to measure the photon arrival times and stream the resulting timestamps to a computer for data analysis.

\section{Spontaneous Raman Scattering in Coexistence Links}

In a shared fiber of length $L$, a classical channel with power $P_C$ and wavelength $\lambda_c$ generates SpRS noise at rate~\cite{eraerds2010quantum, wu2025integration}:
\begin{equation}
    N_R(\lambda) = \frac{P_C}{h\nu_c}\cdot L_{eff}\cdot g_R(\Delta\omega)\cdot\Delta\nu\cdot \eta_{\rm det},
    \label{eq:raman_power}
\end{equation}
where $L_{\rm eff} \approx L\cdot e^{-\alpha_c\cdot L}$, is the effective fiber length with $\alpha_c$ attenuation coefficient, $g_R(\Delta\omega)$ is the Raman gain cross-section at frequency detuning $\Delta\omega$ from the pump, $\Delta\nu$ is the filter bandwidth, and $\eta_{\rm det}$ is the detectors' efficiency. The form of $L_{\rm eff}$ assumes that the attenuation coefficient is the same for both the classical pump and the Raman photons, since the Raman photons are also attenuated when they travel along the remaining length of fiber; for the detailed derivation, check refs. \cite{patel2012coexistence,singh2025photonic}.

The Raman gain profile of silica is exceptionally broad, with its main peak at $\approx 13.2$~THz offset from the pump and significant intensity spanning the entire C-band~\cite{agrawal2012nonlinear}. Regarding other sources of noise, as already mentioned, Brillouin scattering is confined within $\sim$11~GHz and is removed by DWDM filters~\cite{agrawal2012nonlinear}; four-wave mixing requires phase matching and is negligible for a single classical pump~\cite{agrawal2012nonlinear}. SpRS is therefore the only relevant noise source in our configuration.

Accidental coincidences from SpRS raise the noise term in Eq. \eqref{eq:rho_AB}, degrading state visibility when measured with a Franson setup. Once $P_C$ exceeds a certain threshold, the resulting state can no longer generate a secure key for QKD purposes when relying only on classical error correction and privacy amplification. Our distillation protocol absorbs this excess noise into frequency-mode filtering, extending the operational range of $P_C$. The key observation is that SPDC photons always satisfy energy conservation ($\omega_k + \omega_{-k} = \omega_p$), so they are deterministically routed to conjugate channel pairs. Raman noise photons, on the contrary, are uncorrelated and populate all channel combinations with equal probability. Without the proposed frequency-mode post-selection, when signals from all $N$ channels are combined, accidental coincidences scale as $B_{broad}=N^2 B_0$, where $B_0 = R_A R_B\,\Delta t_{\rm coinc}$ is the accidental coincidence rate per channel pair, $R_{A,B}$ being the single-channel noise rates, and $\Delta t_{\rm coinc}$ the set coincidence window. After post-selecting on frequency-matched pairs, the accepted combinations are reduced to $N$, giving:
\begin{equation}
	B_{\rm distilled} = N B_0 = \frac{1}{N}B_{\rm broad}.
\end{equation}
The signal rate $S$ is unchanged, because all genuine pairs already satisfy the energy-conservation condition. The resulting output visibility $V_{\rm dist}$ is therefore:
\begin{equation}
	V_{\rm dist} = \frac{V}{V + \frac{1-V}{N}},
	\label{eq:vis_distilled}
\end{equation}
where $V = S/(S+B_{\rm broad})$ is the non-distilled visibility. The expression is directly derived by writing $V_{\rm dist} = S/(S+B_{\rm broad}/N)$ and dividing numerator and denominator by the total rate $S+B_{\rm broad}$. Equation~(\ref{eq:vis_distilled}) shows that $V_{\rm out} > V$ for any $N > 1$, with $V_{\rm out}\to 1$ as $N\to\infty$. The corresponding post-distillation QBER is:
\begin{equation}
	Q{\rm dist} = \frac{1-V_{\rm dist}}{2} = \frac{1-V}{2[NV + (1-V)]},
\end{equation}
which can fall below the $11\%$ threshold even when the non-distilled QBER $Q=(1-V)/2$, corresponding to the case where all $N$ channels are combined, reducing Eq.~\eqref{eq:hyper_state} to Eq.~\eqref{eq:franson}, exceeds it, restoring secure key generation.

\section{Results}
\begin{figure}[t]
	\centering
	\includegraphics[width=\linewidth]{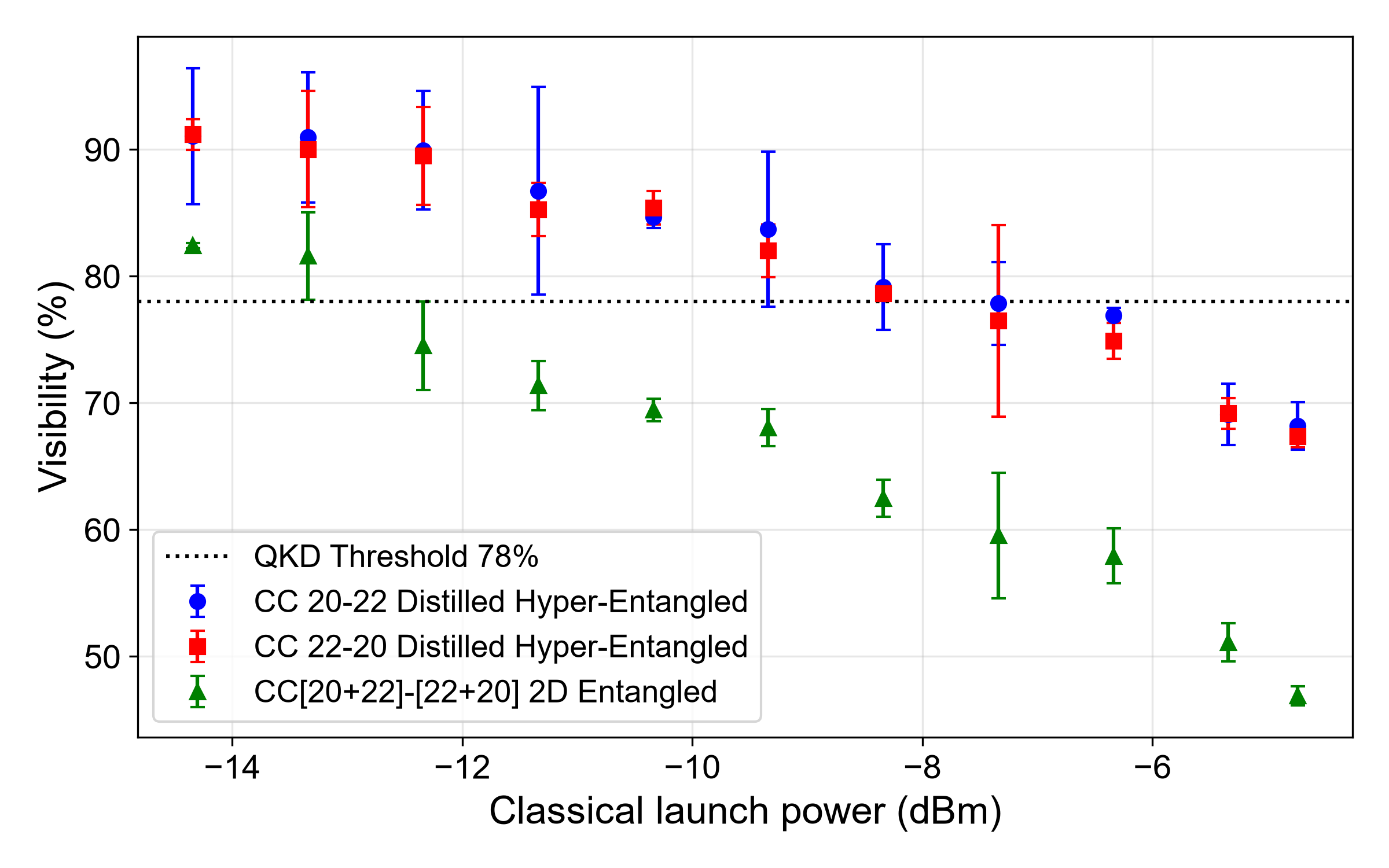}
	\caption{\textbf{Interference visibility versus classical launch power.} Distilled channel pairs (blue, red) maintain visibility above the 78\% QKD security threshold up to \SI{-9.3}{\dBm}, whereas the un-distilled 200-GHz broadband configuration (green) fails at \SI{-13.3}{\dBm}. Error bars represent 1-$\sigma$ uncertainties on $V=|A|/B$ evaluated from the covariance matrix of the sinusoidal fits on Poissonian counting statistics. Point-to-point variations in uncertainty magnitude mainly reflect differences in fringe sampling density.}
	\label{fig:vis}
\end{figure}

In a Franson interferometric scheme, the two two-photon paths $\ket{SS}$ and $\ket{LL}$ interfere according to:
\begin{equation}
	C(\phi_A, \phi_B) \propto 1 + V_0 \cos(\phi_A + \phi_B);
	\label{eq:franson_interference}
\end{equation}
here, $C(\phi_A, \phi_B)$ is the central-peak coincidence rate up to an overall proportionality constant set by the source brightness and detection efficiencies, $\phi_A$ and $\phi_B$ are the phases imparted by Alice's and Bob's iMZIs, respectively, and $V_0$ is the maximum interference visibility attainable in the absence of SpRS noise, intrinsically limited by imperfections in the entangled-state generation. Basis selection is set by Alice's phase offset: the $X$ basis corresponds to $\phi_A = 0$, while the $Y$ basis corresponds to $\phi_A = \pi/2$, set via controlling the temperature of Alice's iMZI. 
The $Z$ basis could be used as the second mutually unbiased basis through time-of-flight measurements~\cite{pelet2023operational}. However, this approach would require additional beam splitters and duplicate single-photon detectors. We instead use the $Y$ basis, which can be measured simply by applying this phase offset.

To characterize the two-photon interference visibility $V$ at each classical launch power and for each basis, Alice's phase $\phi_A$ is fixed at the corresponding basis setting, while Bob's phase $\phi_B$ is swept over a full $2\pi$ period by scanning the temperature T of Bob's iMZI. The resulting coincidence fringe pattern is fitted with
\begin{equation}
	f(T) = A \cos(\omega T + \phi_0) + B,
	\label{eq:fringe_fit}
\end{equation}
from which the visibility is derived as:
\begin{equation}
	V = \frac{A}{B} = \frac{C_{\text{max}} - C_{\text{min}}}{C_{\text{max}} + C_{\text{min}}},
	\label{eq:visibility}
\end{equation}
with ${C_{\text{max}}}$ and ${C_{\text{min}}}$ the maximum and minimum of coincidences, respectively.
We experimentally verify that the visibility measured in the $Y$ basis matches that of the $X$ basis within statistical (Poissonian) fluctuations ($V_Y \approx V_X$), confirming symmetric error rates $E_X = E_Y = (1-V)/2$.

Figure~\ref{fig:vis} shows the two-photon interference visibility as a function of classical launch power. In the absence of classical light, both configurations reach $V_0 \approx 98\%$, limited by detector dark counts, residual PICs imbalances, and, to a lesser extent, state preparation. As $P_C$ increases, SpRS photons accumulate in all channels, raising accidental coincidences and reducing the visibility. The degradation is significantly faster in the broadband scenario, achieved by the two parties by recombining the photons at the output of the DWDM demultiplexers, thereby discarding the frequency degree of freedom and losing the associated entanglement. This can be implemented either physically, using a beam combiner, or, as in our case, by merging the timestamps recorded by their time-tagger channels. The 200-GHz reference then crosses the BBM92 security threshold ($V = 78\%$) at $P_C = -13.3$~dBm, while both distilled channel pairs remain above threshold up to $-9.3$~dBm. The gap between the two curves grows monotonically with power, in quantitative agreement with the $1/N$ suppression predicted by Eq.~(\ref{eq:vis_distilled}).

\begin{figure}[t]
    \centering
    \includegraphics[width=\linewidth]{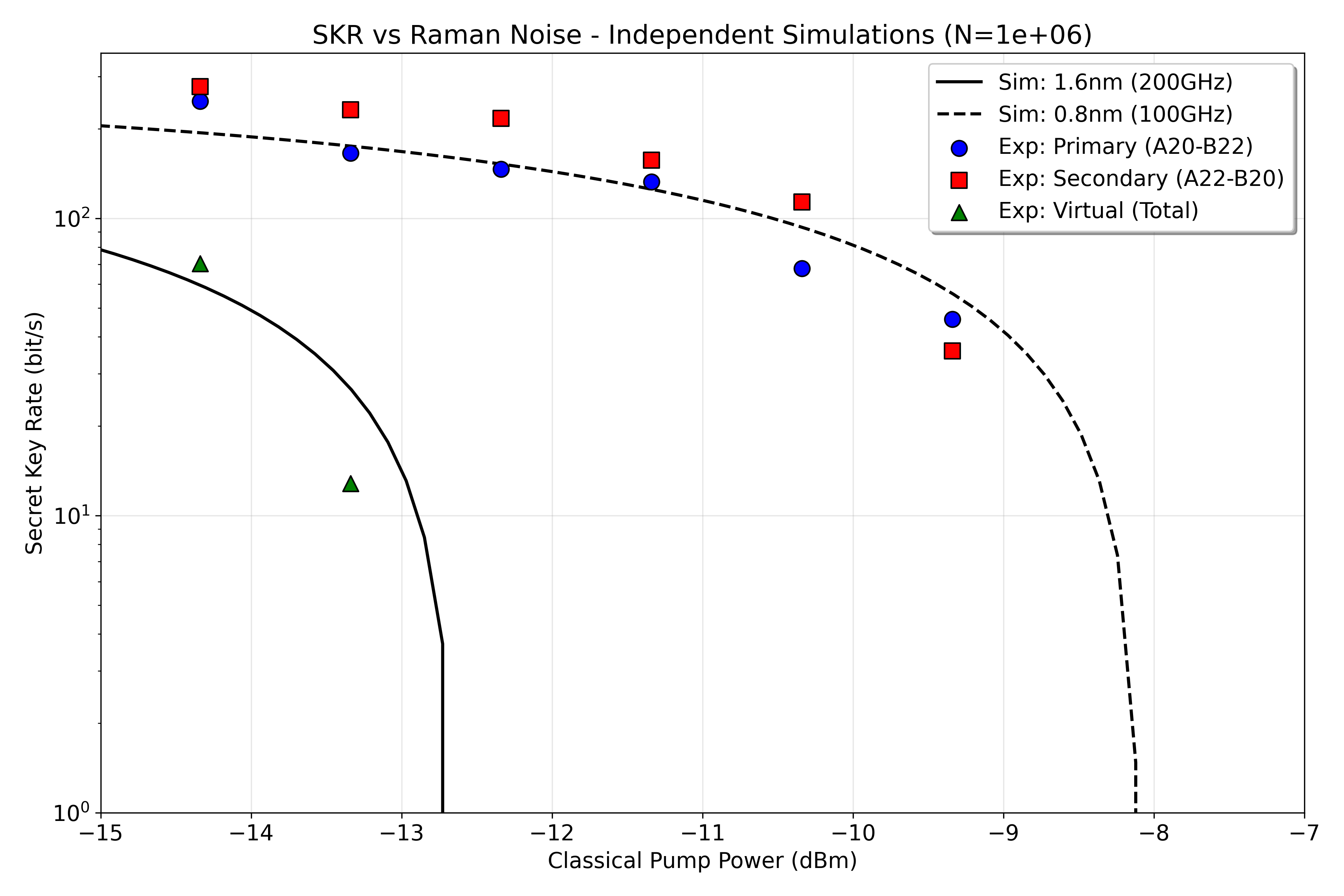}
    \caption{\textbf{Model validation.} Experimental SKR data (markers) versus numerical simulation (solid and dashed lines). Finite-size analysis following Ref.~\cite{tomamichel2012tight} with $f_{\rm EC}=1.15$, $\epsilon_{\rm sec}=10^{-12}$, $\epsilon_{\rm cor}=10^{-12}$, and total sifted block size $N_{\rm tot}=10^6$~bits. Distillation extends the secure key threshold by 4~dB.}
    \label{fig:skr}
\end{figure}

To quantify the practical impact for QKD, we compute the finite-key SKR within the composable framework of Tomamichel \textit{et al.}~\cite{tomamichel2012tight} (see Appendix~\ref{app:skr} for the full derivation). Results are shown in Fig.~\ref{fig:skr}: using the basis error rates $E_X, E_Y$ defined via Eq.~\eqref{eq:visibility}, a block size $N_{\rm tot}=10^6$ and security parameters $\epsilon_{\rm sec}=\epsilon_{\rm cor}=10^{-12}$. The distilled channels sustain a positive key rate up to $P_C = -9.3$~dBm, compared to $-13.3$~dBm for the broadband reference. This means a net gain of 4~dB in classical launch power tolerance, corresponding to a $\approx 2.5\times$ increase in tolerable classical power, or $\approx 20$~km of extended coexistence distance with no hardware changes on the quantum receiver.
We stress that these absolute SKR values, while correctly computed, should not be interpreted as an upper bound: the SKR also depends on the source brightness, which was chosen here for diagnostic purposes rather than to maximize key generation. Specifically, we pumped the source up to a coincidence-to-accidental ratio (CAR) of $\approx 1000$, ensuring that SpRS remained the dominant noise contribution compared to the accidental coincidences arising from multi-pair SPDC emission. In a practical deployment, the pair-generation rate could be increased according to the channel conditions, further raising the achievable SKR.

To corroborate this interpretation, we developed a numerical model of the SpRS noise contribution (Fig.~\ref{fig:skr}). The Raman gain spectrum $g_R(\Delta\omega)$ is digitized from Fig. 1 of Ref.~\cite{eraerds2010quantum} by using the open source software Engauge Digitizer and integrated over the 100-GHz passband of each quantum channel, yielding the expected per-channel noise rate $N_R$ as a function of $P_C$ via Eq.~\eqref{eq:raman_power}. Starting from the measured clean-channel visibility $V_0 \approx 98\%$, the model injects $N_R$ into the coincidence statistics and computes the predicted SKR for both the distilled and broadband scenarios. The agreement with the experimental data across the full power range confirms that the 4-dB gain is well accounted for by the $1/N$ reduction in accidental coincidences.

\section{Conclusion}

Table~\ref{tab:comparison} places our scheme in the context of related approaches. Two-copy distillation protocols~\cite{pan2001experimental, bennett1996purification} require probabilistic two-photon interactions and transmit two independent pairs, scaling poorly with channel loss. Spatial single-copy schemes~\cite{hu2021long} achieve near-unity yield but rely on multi-core fiber and polarization, which is unstable on a fiber channel, limiting their compatibility with installed infrastructure. Ref.~\cite{ecker2021experimental} presents an approach similar to ours; however, it is not easily integrable into standard fiber networks due to its reliance on free-space setups, polarization sensitivity, and the necessity of phase locking. Furthermore, while their noise resilience was benchmarked using a sequence of half-wave plates to degrade polarization entanglement, our work addresses the more complex dynamics of real Raman noise generated within a fiber transmission channel. High-dimensional QKD~\cite{zahidy2024practical, zhong2024hyperentanglement,ecker2019overcoming} intrinsically tolerates higher noise at the cost of increased hardware complexity at both ends, additional detectors, and often non-standard transmission channels. 

\begin{table*}[t]
    \centering
    \small
    \caption{Comparison of noise-mitigation and distillation approaches for entanglement-based QKD.}
    \label{tab:comparison}
    \resizebox{\textwidth}{!}{%
        \begin{tabular}{@{}lcccc@{}}
            \toprule
            \textbf{Approach} & \textbf{Physical resource} & \textbf{Receiver complexity} & \textbf{Protocol yield} & \textbf{Fiber compatibility} \\
            \midrule
            Two-copy distillation~\cite{pan2001experimental} & 2 photon pairs & Probabilistic 2-photon CNOT & $\le 0.5$, scales as $\eta^4$ & SMF \\
            Spatial single-copy~\cite{hu2021long} & Polarization + Path & MCF, polarizing BSs & $\approx 1$, scales as $\eta^2$ & MCF only \\
            Single-copy distillation~\cite{ecker2021experimental} & Polarization + Path & Free-space Franson, PID stabilization & $\approx 1$, scales as $\eta^2$ & With limitations \\
            High-dimensional QKD~\cite{zahidy2024practical, zhong2024hyperentanglement,ecker2019overcoming} & $d$-dim qudits & Multi-port couplers, $d$ detectors & --- & With limitations \\
            MCF space multiplexing~\cite{wu2025integration} & Spatial modes & Standard qubit receiver & --- & MCF only \\
            \textbf{This work} & Time-energy + Frequency  & Passive PIC, standard DWDM & $\approx 1$, scales as $\eta^2$ & \textbf{Telecom-ready} \\
            \bottomrule
        \end{tabular}%
    }
    \vspace{2pt}
\end{table*}

In this work, we demonstrate a practical implementation of QPA in the context of fiber coexistence. Using passive, integrated iMZIs alongside off-the-shelf DWDM components to harness time-energy and frequency hyperentanglement, our scheme exploits the intrinsic frequency anti-correlations of SPDC photons to provide a 4~dB enhancement in classical launch power tolerance over a 20~km standard single-mode fiber channel..

The noise suppression scales as $1/N$ with the number of frequency channel pairs, so fine-grid DWDMs or integrated ring-resonator filters could push this advantage well beyond 4~dB. Furthermore, our method is compatible with other established coexistence techniques -- such as optimized wavelength allocation, time-multiplexed gating, or narrowband spectral filtering -- and could be combined with them to further boost the performances of the system. Reframing DWDM demultiplexers,  traditionally used solely for wavelength routing, as tools for enabling QPA, opens a path toward Raman-resilient quantum networking over the existing telecom infrastructure.

\section*{Acknowledgements}
The authors acknowledge support from the European Research Council under Starting Grant QOMUNE (grant agreement No.~101077917).

\section*{Disclosures}
The authors declare no competing interests.

\section*{Data availability}
Data supporting the findings of this study are available from the corresponding author upon reasonable request.

\section{Methods}
\subsection{Finite-key Secure Key Rate}
\label{app:skr}

In an entanglement-based BBM92 protocol, Alice and Bob accumulate a total block of $N_{\rm tot}$ sifted coincidences over an acquisition time $T_{\rm fill}=N_{\rm tot}/R_{\rm sift}$. Here $R_{\rm sift}=(B+|A|)/2$ is the sifted coincidence rate, with $A$ and $B$ as defined in Eq.~\eqref{eq:fringe_fit}. Assuming a symmetric basis choice via passive 50:50 beam splitters, the sifted block is partitioned into $n_X = N_{\rm tot}/2$ key-generation bits (measured in the $X$ basis) and $k_Y = N_{\rm tot}/2$ test bits used to bound the phase error rate in the conjugate basis $Y$.

Due to finite sampling statistics, the phase error rate of the unmeasured key block deviates from the observed test-basis error rate $E_Y$. Applying Serfling's inequality for sampling without replacement~\cite{tomamichel2012tight}, the phase error rate is upper-bounded with failure probability $\epsilon_{\rm sec}$ by
\begin{equation}
	E_Y^{\rm UB} = E_Y + \mu(n_X, k_Y, \epsilon_{\rm sec}),
\end{equation}
with
\begin{equation}
	\mu(n_X, k_Y, \epsilon_{\rm sec}) = \sqrt{\frac{n_X + k_Y}{n_X k_Y} \frac{k_Y + 1}{k_Y} \ln\left(\frac{4}{\epsilon_{\rm sec}}\right)}.
\end{equation}
The extractable secret key length $\ell$, guaranteeing $\epsilon_{\rm sec}$-secrecy and $\epsilon_{\rm cor}$-correctness, is~\cite{tomamichel2012tight}
\begin{equation}
	\begin{aligned}
		\ell = \max\bigg(0, \; & n_X \left[1 - f_{\rm EC}\,h(E_X) - h(E_Y^{\rm UB})\right] \\
		& - \log_2\frac{2}{\epsilon_{\rm sec}^2 \epsilon_{\rm cor}}\bigg),
	\end{aligned}
	\label{eq:skr_finite}
\end{equation}
where $h(p) = -p\log_2 p - (1-p)\log_2(1-p)$ is the binary Shannon entropy, $f_{\rm EC}=1.15$ is the error-correction efficiency, and the logarithmic term accounts for privacy amplification and error-verification leakage. The resulting secret key rate (bit/s) is
\begin{equation}
	R_{\rm SKR} = \frac{\ell}{T_{\rm fill}} = \frac{\ell}{N_{\rm tot}} R_{\rm sift}.
\end{equation}

\bibliographystyle{unsrt}
\bibliography{references}

@article{cozzolino2019high,
  title={High-dimensional quantum communication: benefits, progress, and future challenges},
  author={Cozzolino, Daniele and Da Lio, Beatrice and Bacco, Davide and Oxenl{\o}we, Leif Katsuo},
  journal={Advanced Quantum Technologies},
  volume={2},
  number={12},
  pages={1900038},
  year={2019},
  publisher={Wiley Online Library}
}

@article{wu2025integration,
  title={Integration of quantum key distribution and high-throughput classical communications in field-deployed multi-core fibers},
  author={Wu, Qi and Ribezzo, Domenico and Di Sciullo, Giammarco and Cocchi, Sebastiano and Ann Shaji, Divya and Alves Zischler, Lucas and Luis, Ruben and Serena, Paolo and Lasagni, Chiara and Bononi, Alberto and others},
  journal={Light: Science \& Applications},
  volume={14},
  number={1},
  pages={274},
  year={2025},
  publisher={Nature Publishing Group UK London}
}

@article{guarda2024quantum,
  title={Quantum key distribution with a borosilicate-glass-matrix integrated photonic receiver},
  author={Guarda, Giulia and Ribezzo, Domenico and Occhipinti, Tommaso and Zavatta, Alessandro and Bacco, Davide},
  journal={Physical Review A},
  volume={110},
  number={4},
  pages={042605},
  year={2024},
  publisher={APS}
}

@article{franson1989bell,
  title={Bell inequality for position and time},
  author={Franson, James D},
  journal={Physical Review Letters},
  volume={62},
  number={19},
  pages={2205},
  year={1989},
  publisher={APS}
}

@article{ecker2021experimental,
  title={Experimental single-copy entanglement distillation},
  author={Ecker, Sebastian and Sohr, Philipp and Bulla, Lukas and Huber, Marcus and Bohmann, Martin and Ursin, Rupert},
  journal={Physical Review Letters},
  volume={127},
  number={4},
  pages={040506},
  year={2021},
  publisher={APS}
}

@article{sohr2024overcoming,
  title={Overcoming noise limitations in quantum key distribution with quantum privacy amplification},
  author={Sohr, Philipp and Ecker, Sebastian and Bulla, Lukas and Bohmann, Martin and Ursin, Rupert},
  journal={Physical Review Applied},
  volume={22},
  number={2},
  pages={024059},
  year={2024},
  publisher={APS}
}

@article{hu2021long,
  title={Long-distance entanglement purification for quantum communication},
  author={Hu, Xiao-Min and Huang, Cen-Xiao and Sheng, Yu-Bo and Zhou, Lan and Liu, Bi-Heng and Guo, Yu and Zhang, Chao and Xing, Wen-Bo and Huang, Yun-Feng and Li, Chuan-Feng and Guo, Guang-Can},
  journal={Physical Review Letters},
  volume={126},
  number={1},
  pages={010503},
  year={2021},
  publisher={APS}
}

@article{zhong2024hyperentanglement,
  title={Hyperentanglement quantum communication over a 50 km noisy fiber channel},
  author={Zhong, Zhen-Qiu and Zhan, Xiao-Hai and Chen, Jia-Lin and Wang, Shuang and Yin, Zhen-Qiang and Geng, Jia-Qi and He, De-Yong and Chen, Wei and Guo, Guang-Can and Han, Zheng-Fu},
  journal={Optica},
  volume={11},
  number={8},
  pages={1056--1061},
  year={2024},
  publisher={Optica Publishing Group}
}

@article{pelet2023operational,
  title={Operational entanglement-based quantum key distribution over 50 km of field-deployed optical fibers},
  author={Pelet, Yoann and Sauder, Gr{\'e}gory and Cohen, Mathis and Labont{\'e}, Laurent and Alibart, Olivier and Martin, Anthony and Tanzilli, S{\'e}bastien},
  journal={Physical Review Applied},
  volume={20},
  number={4},
  pages={044006},
  year={2023},
  publisher={APS}
}

@article{wehner2018quantum,
  title={Quantum internet: A vision for the road ahead},
  author={Wehner, Stephanie and Elkouss, David and Hanson, Ronald},
  journal={Science},
  volume={362},
  number={6412},
  pages={eaam9288},
  year={2018},
  publisher={American Association for the Advancement of Science}
}

@article{zahidy2024practical,
  title={Practical high-dimensional quantum key distribution protocol over deployed multicore fiber},
  author={Zahidy, Mujtaba and Ribezzo, Domenico and De Lazzari, Claudia and Vagniluca, Ilaria and Biagi, Nicola and M{\"u}ller, Ronny and Occhipinti, Tommaso and Oxenl{\o}we, Leif K and Galili, Michael and Hayashi, Tetsuya and others},
  journal={Nature Communications},
  volume={15},
  number={1},
  pages={1651},
  year={2024},
  publisher={Nature Publishing Group UK London}
}

@article{simon2002polarization,
  title={Polarization entanglement purification using spatial entanglement},
  author={Simon, Christoph and Pan, Jian-Wei},
  journal={Physical Review Letters},
  volume={89},
  number={25},
  pages={257901},
  year={2002},
  publisher={APS}
}

@article{barreiro2005generation,
  title={Generation of hyperentangled photon pairs},
  author={Barreiro, Julio T and Langford, Nathan K and Peters, Nicholas A and Kwiat, Paul G},
  journal={Physical Review Letters},
  volume={95},
  number={26},
  pages={260501},
  year={2005},
  publisher={APS}
}

@article{deutsch1996quantum,
  title={Quantum privacy amplification and the security of quantum cryptography over noisy channels},
  author={Deutsch, David and Ekert, Artur and Jozsa, Richard and Macchiavello, Chiara and Popescu, Sandu and Sanpera, Anna},
  journal={Physical Review Letters},
  volume={77},
  number={13},
  pages={2818},
  year={1996},
  publisher={APS}
}

@article{bennett1996purification,
  title={Purification of noisy entanglement and faithful teleportation via noisy channels},
  author={Bennett, Charles H and Brassard, Gilles and Popescu, Sandu and Schumacher, Benjamin and Smolin, John A and Wootters, William K},
  journal={Physical Review Letters},
  volume={76},
  number={5},
  pages={722},
  year={1996},
  publisher={APS}
}

@article{shor2000simple,
  title={Simple proof of security of the BB84 quantum key distribution protocol},
  author={Shor, Peter W and Preskill, John},
  journal={Physical Review Letters},
  volume={85},
  number={2},
  pages={441},
  year={2000},
  publisher={APS}
}

@article{kwiat2001experimental,
  title={Experimental entanglement concentration and knotting using linear optics},
  author={Kwiat, Paul G and Barraza-Lopez, Salvador and Stefanov, Andr{\'e} and Gisin, Nicolas},
  journal={Nature},
  volume={409},
  number={6823},
  pages={1014--1017},
  year={2001},
  publisher={Nature Publishing Group}
}

@article{kumar2015coexistence,
  title={Coexistence of continuous variable QKD with intense DWDM classical channels},
  author={Kumar, Rupesh and Qin, Hao and All{\'e}aume, Romain},
  journal={New Journal of Physics},
  volume={17},
  number={4},
  pages={043027},
  year={2015},
  publisher={IOP Publishing}
}

@article{patel2012coexistence,
  title={Coexistence of high-bit-rate quantum key distribution and data on optical fiber},
  author={Patel, KA and Dynes, JF and Choi, I and Sharpe, AW and Dixon, AR and Yuan, ZL and Penty, RV and Shields, AJ},
  journal={Physical Review X},
  volume={2},
  number={4},
  pages={041010},
  year={2012},
  publisher={APS}
}

@article{eraerds2010quantum,
  title={Quantum key distribution and 1 Gbps data encryption over a single fibre},
  author={Eraerds, Patrick and Walenta, Nino and Legr{\'e}, Matthieu and Gisin, Nicolas and Zbinden, Hugo},
  journal={New Journal of Physics},
  volume={12},
  number={6},
  pages={063027},
  year={2010},
  publisher={IOP Publishing}
}

@book{agrawal2012nonlinear,
  title={Nonlinear Fiber Optics},
  author={Agrawal, Govind P},
  year={2012},
  publisher={Academic Press},
  edition={5th}
}

@article{pan2001experimental,
  title={Entanglement purification for quantum communication},
  author={Pan, Jian-Wei and Simon, Christoph and Brukner, {\v{C}}aslav and Zeilinger, Anton},
  journal={Nature},
  volume={410},
  number={6832},
  pages={1067--1070},
  year={2001},
  publisher={Nature Publishing Group}
}

@article{tomamichel2012tight,
  title={Tight finite-key analysis for quantum cryptography},
  author={Tomamichel, Marco and Lim, Charles Ci Wen and Gisin, Nicolas and Renner, Renato},
  journal={Nature Communications},
  volume={3},
  number={1},
  pages={634},
  year={2012},
  publisher={Nature Publishing Group UK London}
}

@article{o2003demonstration,
	title={Demonstration of an all-optical quantum controlled-NOT gate},
	author={O'Brien, Jeremy L and Pryde, Geoffrey J and White, Andrew G and Ralph, Timothy C and Branning, David},
	journal={Nature},
	volume={426},
	number={6964},
	pages={264--267},
	year={2003},
	publisher={Nature Publishing Group UK London}
}

@article{gasparoni2004,
	title={Realization of a photonic controlled-not gate sufficient for quantum computation},
	url={https://journals.aps.org/prl/abstract/10.1103/PhysRevLett.93.020504},
	journal={Physical Review Letters},
	publisher={American Physical Society},
	author={Gasparoni, Sara and Pan, Jian-Wei and Walther, Philip and Rudolph, Terry and Zeilinger, Anton},
	year={2004},
	month={Jul}
}

@article{horodecki2001distillation,
	title={Distillation and bound entanglement.},
	author={Horodecki, Pawel and Horodecki, Ryszard},
	journal={Quantum Inf. Comput.},
	volume={1},
	number={1},
	pages={45--75},
	year={2001}
}

@article{singh2025photonic,
	title={Photonic quantum information with time-bins: Principles and applications},
	author={Singh, Ashutosh and Sethia, Anuj and Esmaeilifar, Leili and Valivarthi, Raju and Sinclair, Neil and Spiropulu, Maria and Oblak, Daniel},
	journal={arXiv preprint arXiv:2507.08102},
	year={2025}
}

@article{shapiro2006single,
  title={Single-photon Kerr nonlinearities do not help quantum computation},
  author={Shapiro, Jeffrey H},
  journal={Physical Review A—Atomic, Molecular, and Optical Physics},
  volume={73},
  number={6},
  pages={062305},
  year={2006},
  publisher={APS}
}

@article{wegener2004effect,
  title={The effect of propagation nonlinearities on the information capacity of WDM optical fiber systems: Cross-phase modulation and four-wave mixing},
  author={Wegener, LGL and Povinelli, BML and Green, AG and Mitra, PP and Stark, JB and Littlewood, PB},
  journal={Physica D: Nonlinear Phenomena},
  volume={189},
  number={1-2},
  pages={81--99},
  year={2004},
  publisher={Elsevier}
}

@article{bennett1996concentrating,
  title={Concentrating partial entanglement by local operations},
  author={Bennett, Charles H and Bernstein, Herbert J and Popescu, Sandu and Schumacher, Benjamin},
  journal={Physical Review A},
  volume={53},
  number={4},
  pages={2046},
  year={1996},
  publisher={APS}
}

@article{ecker2019overcoming,
  title={Overcoming noise in entanglement distribution},
  author={Ecker, Sebastian and Bouchard, Fr{\'e}d{\'e}ric and Bulla, Lukas and Brandt, Florian and Kohout, Oskar and Steinlechner, Fabian and Fickler, Robert and Malik, Mehul and Guryanova, Yelena and Ursin, Rupert and others},
  journal={Physical Review X},
  volume={9},
  number={4},
  pages={041042},
  year={2019},
  publisher={APS}
}

\end{document}